\documentclass[fleqn,usenatbib]{mnras}

\usepackage{newtxtext,newtxmath}

\usepackage[T1]{fontenc}

\DeclareRobustCommand{\VAN}[3]{#2}
\let\VANthebibliography\thebibliography
\def\thebibliography{\DeclareRobustCommand{\VAN}[3]{##3}\VANthebibliography}

\usepackage{graphicx}	
\usepackage{amsmath}	
\usepackage{orcidlink}

\newcommand{\mesa}{{\tt MESA}}
\newcommand{\Msun}{\ensuremath{{\rm M}_\odot}}

\title[Slowest-spinning field spider: PSR\,J1932+2121]{The slowest spinning Galactic-field spider PSR\,J1932+2121: A history of inefficient mass transfer}

\author[Misra et al.]{
Devina Misra,$^{\orcidlink{0000-0003-4260-960X}}$$^{1}$\thanks{E-mail: devina.misra@ntnu.no},
Karri I. I. Koljonen$^{\orcidlink{0000-0002-9677-1533}}$$^{1}$,
Manuel Linares$^{\orcidlink{0000-0002-0237-1636}}$ $^{1,2}$
\\
$^{1}$Department of Physics, Norwegian University of Science and Technology, NO-7491 Trondheim, Norway,\\
$^{2}$Departament de F{\'i}sica, EEBE, Universitat Polit{\`e}cnica de Catalunya, Av. Eduard Maristany 16, E-08019 Barcelona, Spain
}

\date{Accepted XXX. Received YYY; in original form ZZZ}

\pubyear{\the\year{}}

\begin{document}
\label{firstpage}
\pagerange{\pageref{firstpage}--\pageref{lastpage}}
\maketitle

\begin{abstract}
The Five-hundred-meter Aperture Spherical Telescope is discovering hundreds of new pulsars, including a slowly spinning compact binary millisecond pulsar (spin period $P_{\rm spin}=14.2$\,ms) which showed radio eclipses and evidence of ablation of its companion: PSR J1932+2121. Its orbital period is $P_{\rm orb}=0.08$\,d and the minimum companion mass is estimated as 0.12\,\Msun. Hence, this pulsar is classified as part of the Galactic-field spider (redback) population. However, it spins almost an order of magnitude slower than other Galactic-field spiders. 
Using detailed evolutionary calculations with \mesa, we model the formation, mass-transfer and radio-pulsar phases, in order to explain the observed properties of PSR\,J1932+2121.
We find that PSR\,J1932+2121 is a redback that has experienced an inefficient mass-transfer phase resulting in a lower accretion efficiency (in the range of 0.3 to 0.5) and subsequently slower spin compared to other spiders. We narrow down the initial range of $P_{\rm orb}$ that best reproduces its properties, to 2.0--2.6\,d. Current models of accretion-induced magnetic field decay are not able to explain its unusually high surface magnetic field of $2\times 10^{9}$\,G.
Hence, PSR\,J1932+2121 provides a unique opportunity to study inefficient accretion-induced spin up and surface magnetic field decay of pulsars.
\end{abstract}




\section{Introduction}
Compact millisecond pulsars (CBMPs; with spins $\lesssim30$\,ms and orbital periods $\lesssim1$\,d) form after a neutron star (NS) has spun up during a mass-transfer phase \citep{1982Natur.300..728A,8812f611-e297-3575-a70c-49607c391972,BHATTACHARYA19911}. Observational studies link accreting low-mass X-ray binaries (LMXBs) to CBMPs via this recycling process \citep{1998Natur.394..344W,2009Sci...324.1411A}. As the NS accretes matter, its surface magnetic field is suppressed due to Ohmic decay \citep{1997MNRAS.284..311K,1998A&A...330..195Z,1999MNRAS.303..588K, 1999MNRAS.308..795K, 2001ApJ...557..958C}, explaining why millisecond pulsars (MSPs) exhibit significantly weaker magnetic fields ($\sim 10^8$\,G) compared to young pulsars ($\sim 10^{12}$\,G). A subset of these CBMPs show eclipses in their radio observations indicating their companions are irradiated by the pulsar \citep{2009Sci...324.1411A}. These eclipsing CBMPs are also called spiders due to the cannibalistic nature of this interaction, as pulsar irradiation gradually erodes the companion star. Spiders are classified into two main groups, based on the masses of the companion stars ($M_{\rm c}$): redbacks (RBs), with companions masses of $0.1{\,\Msun}\lesssim M_{\rm c}\lesssim0.5\,\Msun$ and black widows (BWs), with companions masses of $M_{\rm c}<0.1\,\Msun$ \citep{2013IAUS..291..127R}. More recently, an additional class of spiders, known as tidarrens, has been identified \citep[with orbital periods of $P_{\rm orb}\lesssim 2$\,hrs and $M_{\rm c}\lesssim 0.015\,\Msun$;][]{2016ApJ...833..138R}. 

\citet{2013ApJ...775...27C} proposed that RBs and BWs represent two distinct populations formed under different strengths of pulsar wind irradiation. \citet{2014ApJ...786L...7B,2015ApJ...798...44B} included X-ray irradiation feedback during the Roche-lobe overflow (RLO) phase and suggested that RBs evolve into BWs due to cyclic mass transfer. Recently, \citet{2025A&A...693A.314M} carried out detailed binary simulations, exploring the effects of accretion efficiency, companion evolutionary state and pulsar wind irradiation, on the spin up of pulsars and the formation of spiders. They found that while RBs can evolve into BWs, not all BWs necessarily originate from a RB stage. 

\begin{figure}
\centering
\includegraphics[width=\linewidth]{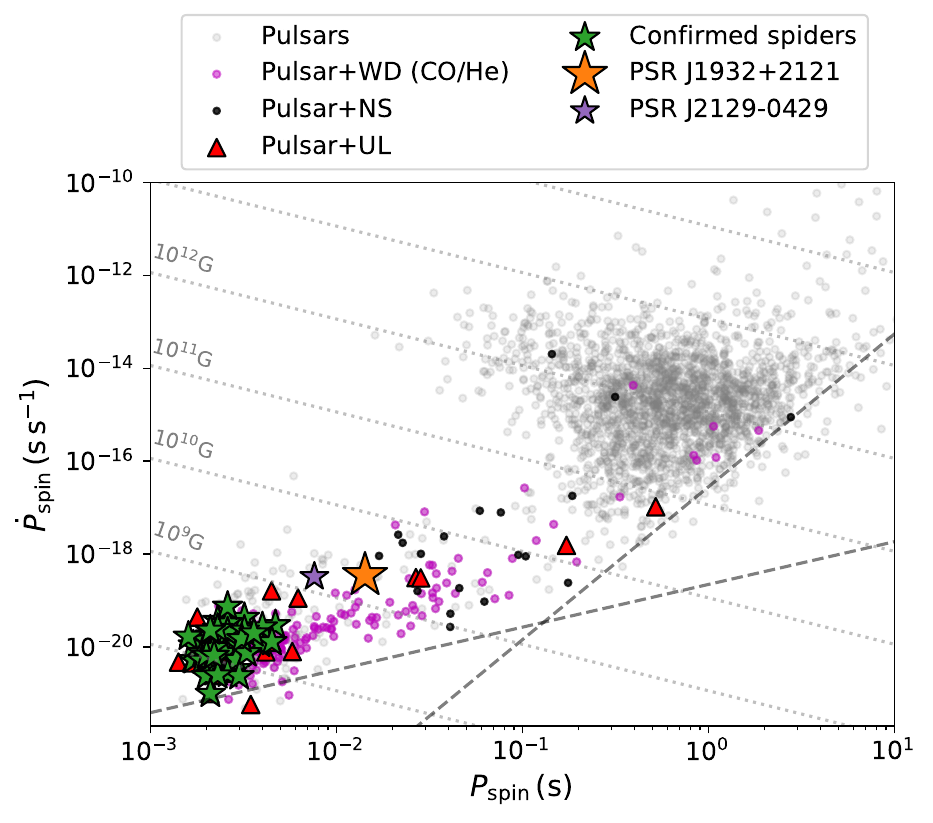}
\caption{{Spin period derivative ($\dot{P}_{\rm spin}$) versus spin period ($P_{\rm spin}$) of observed pulsars. Dotted grey lines of constant magnetic field are indicated. The faint grey circles are pulsar observations taken from ATNF Pulsar Catalog\protect\footnotemark \citep{2005AJ....129.1993M} and the green stars are confirmed spiders \citep[][and references therein]{Nedreaas_master_thesis}. We distinguish between the different binaries present in the Galactic field, pink circles are the pulsar binaries with carbon-oxygen (CO) or helium (He) white dwarfs (WDs), black dots are pulsar-NS binaries and red triangles are pulsars with ultra-light (UL) companions ($M_{\rm c,min}\lesssim0.02$\,\Msun). The orange and purple stars are the $M_{\rm c,min}$ and $P_{\rm orb}$ for PSR\,J1932+2121 \citep{2024arXiv241203062W} and PSR\,J2129-0429 \citep{2024ApJ...966..161B}, respectively. The two dashed grey lines are the pulsar death lines \citep{1994MNRAS.267..513R}.}}
\label{fig:Psdot_Ps_data}
\end{figure}

\citet{2024arXiv241203062W} presented 116 newly discovered pulsars in the Galactic Plane Pulsar Snapshot survey (GPPS) using the Five-hundred-meter Aperture Spherical radio Telescope (FAST). Among these they identified and confirmed two new BWs and three RBs, along with 12 spider candidates. One of the newly discovered RBs, PSR\,J1932+2121, shows peculiar characteristics that differ from other RBs. It has a spin period ($P_{\rm spin}$) of $14.2$\,ms and spin period derivative ($\dot{P}_{\rm spin}$) of $3.53\times 10^{-19}$\,s\,s$^{-1}$. All other observed spiders to date have pulsar spin periods $\lesssim8$\,ms \citep[][and references therein]{Nedreaas_master_thesis} placing PSR\,J1932+2121 in the mildly-recycled regime of MSPs (see Table\,\ref{tab:J1932} for its observed and estimated properties). Its orbital period ($P_{\rm orb}$) of 1.94\,hrs makes it the most compact RB currently known. Figure\,\ref{fig:Psdot_Ps_data} highlights the difference in spin period, spin period derivative and derived surface magnetic field strength, when comparing PSR\,J1932+2121 to other spiders.

Considering that the degree of orbital contraction in LMXBs depends on the amount of mass transferred during the recycling process, it is puzzling how PSR\,J1932+2121 achieved such a compact orbit while failing to spin up to values typical to spiders. Building upon our previous work investigating the spider formation using detailed evolutionary modeling \citep[][]{2025A&A...693A.314M}, we investigate here the possible pathways resulting in the peculiar slow spin and the compact orbit of PSR\,J1932+2121. In Section\,\ref{sec:num_methods}, we describe the numerical code used, the physics of pulsar wind irradiation and accretion-induced magnetic field decay in pulsars. In Section\,\ref{sec:results} we compare our simulated binary tracks to the properties of PSR\,J1932+2121. Section\,\ref{sec:discussion} presents the discussion of our results, followed by conclusions in Section\,\ref{sec:conclusions}.

\begin{table}
\begin{center}
\caption{Observed and estimated properties of PSR\,J1932+2121 \citep{2024arXiv241203062W}. $P_{\rm orb}$ is the orbital period, $P_{\rm spin}$ and $\dot{P}_{\rm spin}$ are the spin period and spin period derivative, $B_{\rm NS}$ is the estimated surface magnetic field, $M_{\rm c, min}$ is the minimum companion mass and $M_{\rm c, med}$ is the median companion mass. Since we know the spin and its derivative we can calculate the spin-down luminosity as $L_{\rm pulsar}$.} 
\begin{tabular}{l|l}
\hline
$P_{\rm orb}$       & 0.0809\,d                   \\
$P_{\rm spin}$      & 14.245\,ms                  \\
$\dot{P}_{\rm spin}$ & $3.53\times 10^{-19}$\,s\,s$^{-1}$ \\
$B_{\rm NS}$      & $2.27\times 10^{9}$\,G        \\
$M_{\rm c, min}$    & 0.115\,\Msun                \\
$M_{\rm c, med}$    & 0.134\,\Msun                \\ 
$L_{\rm pulsar}$    & $4.81\times 10^{33}$\,erg\,s$^{-1}$           \\ \hline
\end{tabular}
\label{tab:J1932}
\end{center}
\end{table}

\footnotetext{https://www.atnf.csiro.au/research/pulsar/psrcat} 

\section{Methods}
\label{sec:num_methods}

To carry out our study, we use the detailed stellar evolution code Modules for Experiments in Stellar Astrophysics
\citep[\mesa;][]{2011ApJS..192....3P,2013ApJS..208....4P,2015ApJS..220...15P,2018ApJS..234...34P,2019ApJS..243...10P}. The observed and estimated parameters for this source are shown in Table\,\ref{tab:J1932}, with which we compare our simulated evolutionary tracks. The general stellar and binary physics follow the description provided by \citet{2023ApJS..264...45F}. We calculate the NS spin evolution and pulsar wind irradiation following the description in \citet{2025A&A...693A.314M}, with relevant details mentioned below. We also compare our simulations with the rest of the spider population \citep[][and references therein]{Nedreaas_master_thesis}. The observations constrain the minimum companion mass (denoted as $M_{\rm c,min}$) calculated assuming a pulsar mass of $1.4\,\Msun$ and inclination of $90^\circ$ from radio observations.

 The efficiency of accretion during RLO determines the final spin and mass of the pulsar. To account for non-conservative mass transfer we consider potential mass lost from the system during the RLO phase. This is parametrized as $\beta$: the fraction of mass transferred that leaves the system from the vicinity of the pulsar as fast isotropic wind. \citet{2025A&A...693A.314M} found that for an X-ray irradiated accretion disc, $\beta\lesssim 0.3$ is required to reproduce the observed spider spins ($\lesssim 8$\,ms). Since PSR\,J1932+2121 spins slower than other spiders (see Figure\,\ref{fig:Psdot_Ps_data}), here we extend the range of $\beta$ values up to 0.7.

As the companion loses mass during RLO, it forms a fully convective structure around 0.2--0.3\,\Msun \citep{1983ApJ...275..713R} and the RLO phase is halted. The pulsar then spins down and emits an energetic wind, which irradiates the outer envelope of the companion and induces mass loss via a stellar wind \citep{1988Natur.334..225K, 1989ApJ...343..292R, 1989ApJ...336..507R, 1996ApJ...473L.119S}. This wind disperses the radio emission from the pulsar causing eclipses in the radio light curves when the wind is located between our line-of-sight and the pulsar. The effect of pulsar wind irradiation is calculated using the spin-down luminosity, which is described as follows:
\begin{equation}\label{eq:spindown_L}
    L_{\rm pulsar} = \frac{4\pi^{2} I \dot{P}_{\rm spin}}{{P}_{\rm spin}^{3}},
\end{equation}
where $I$ is the pulsar moment of inertia, calculated for a solid sphere as $2M_{\rm NS}R_{\rm NS}^{2}/5$, $M_{\rm NS}$ and $R_{\rm NS}$ are the NS mass and radius, respectively. {For $R_{\rm NS}$, we use 12.5\,km \citep{2020PhRvD.101l3007L,2020ApJ...892L...3A,2021ApJ...921...63B,2021ApJ...918L..29R}. Initial values of $P_{\rm spin}$ and $\dot{P}_{\rm spin}$ are taken as $1\,\rm s$ and $10^{-15}\,\rm s/s$ (typical for young pulsars), making the corresponding initial $B_{\rm NS}\sim10^{12}\,\rm G$.} The mass loss from the companion due to pulsar wind irradiation is calculated as follows \citep{1992MNRAS.254P..19S}:
\begin{equation}\label{eq:mdot_irrad}
    \dot{M}_{\rm c, irr} = -f_{\rm pulsar}\times\frac{L_{\rm pulsar}}{2v^{2}_{\rm esc}}\bigg(\frac{R_{\rm c}}{a}\bigg)^{2},
\end{equation}
where $v_{\rm esc}$ is the escape velocity from the companion surface, $R_{\rm c}$ is the companion radius, $a$ is the binary separation and $f_{\rm pulsar}$ is the efficiency of converting spin-down luminosity into kinetic energy of the wind. In the literature, the values explored for $f_{\rm pulsar}$ range from 0.0 to 0.5 \citep{2013ApJ...775...27C,2025A&A...693A.314M}. Since, PSR\,J1932+2121 shows active irradiation as evidenced by the observed radio eclipses \citep{2024arXiv241203062W}, we consider $f_{\rm pulsar}>0.0$. 

The surface magnetic field of a pulsar is thought to be buried as the pulsar accretes matter \citep{1997MNRAS.284..311K, 1999MNRAS.303..588K, 1999MNRAS.308..795K}. This introduces accretion-induced decay in magnetic field 
{\citep{2011MNRAS.413..461O}} which can be expressed as:
\begin{equation}\label{eq:B_accretion}
    B_{\rm NS} = (B^{i} - B_{\rm min})\times\exp{(-\Delta M_{\rm NS}/M_{\rm d})} + B_{\rm min}\,,
\end{equation}
where, $B_{\rm NS}$ is the calculated surface magnetic field of the NS, $B^{i}$ is the initial surface magnetic field of the NS (which we take as $10^{12}$\,G), $\Delta M_{\rm NS}$ is the accreted mass, $M_{\rm d}$ is the magnetic field mass decay scale and $B_{\rm min}$ is the minimum magnetic field \citep[we assume $B_{\rm min}=10^8\,\rm G$;][]{2006MNRAS.366..137Z,2011MNRAS.413..461O}. For $M_{\rm d}$, a range of values has been investigated \citep[0.0033 to 0.05\,\Msun;][]{2011MNRAS.413..461O,2020MNRAS.494.1587C}: we assume an intermediate value of 0.025\,\Msun. Since PSR\,J1932+2121 has a higher estimated surface magnetic field than other spiders ($B_{\rm NS}=2.27\times 10^9$\,G; see Figure\,\ref{fig:Psdot_Ps_data}), we also consider a higher value of $M_{\rm d}=$0.075\,\Msun, which would slow down the magnetic field decay. {The surface magnetic field of the NS affects the angular velocity of the accreted material, which is defined as follows:
\begin{equation}\label{eq:omega}
    \omega_{\rm diff} = \sqrt{\frac{GM_{\rm NS}}{r^{3}_{\rm mag}}} - \sqrt{\frac{GM_{\rm NS}}{r^{3}_{\rm cor}}},
\end{equation} 
where $G$ is the gravitational constant, $r_{\rm cor}$ is the co-rotation radius and $r_{\rm mag}$ is the magnetospheric radius. As the material transfers angular momentum to the NS, a larger $r_{\rm mag}$ would lead to a lesser degree of NS spin up than a smaller $r_{\rm mag}$.
}

\section{Results}
\label{sec:results}

\begin{figure}
\centering
\includegraphics[width=0.9\linewidth]{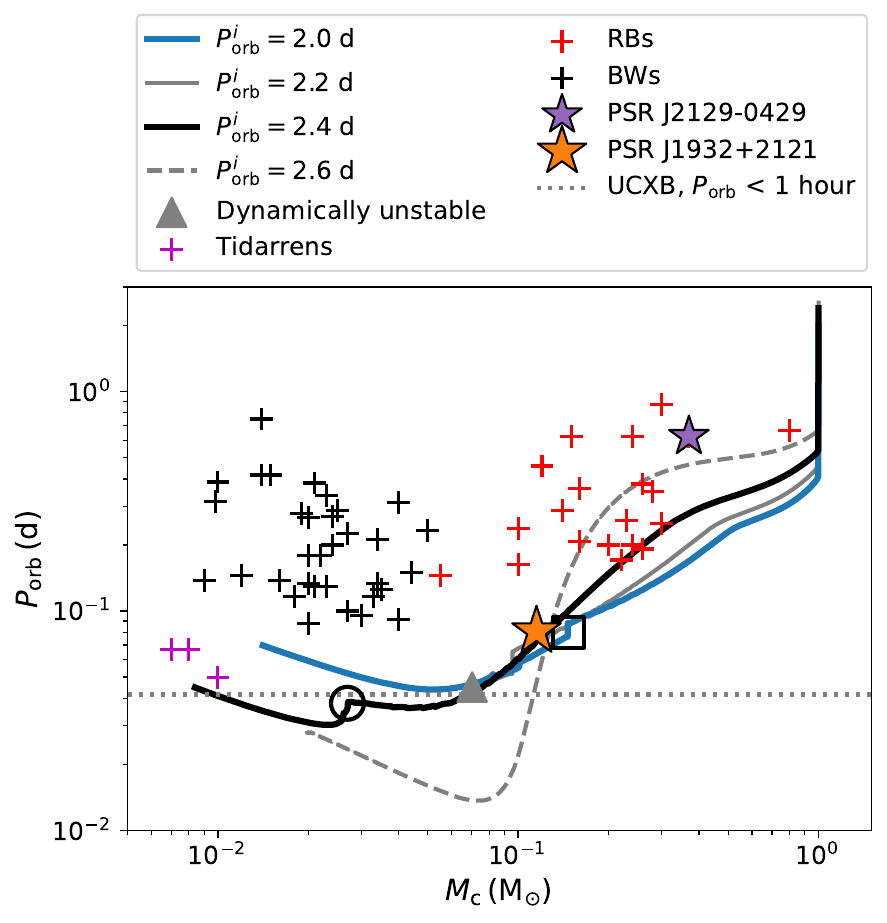}
\caption{Orbital period ($P_{\rm orb}$) evolution versus companion mass ($M_{\rm c}$) of the simulated binaries for $M^{i}_{\rm NS}=1.3\,\Msun$ and $M^{i}_{\rm c}=1.0\,\Msun$, with $\beta=0.7$ and $f_{\rm pulsar}=0.05$. The tracks shown best reproduce the observed orbital parameters of PSR\,J1932+2121 and correspond to $P^{i}_{\rm orb}$ in the range of 2.0 to 2.6\,d (see legend). The track with $P^{i}_{\rm orb}=2.2$\,d ends up being dynamically unstable, shown by the solid grey triangle. The square and circle symbols show when the companion develop a fully convective structure for case\,A ($P^{i}_{\rm orb}=2.0$\,d) and near-$P_{\rm bif}$ ($P^{i}_{\rm orb}=2.4$\,d) RLO cases, respectively. We compare the tracks to estimated $M_{\rm c,min}$ and observed $P_{\rm orb}$ for BWs (black crosses), RBs (red crosses) and tidarrens (purple crosses) from \citet[][]{Nedreaas_master_thesis}. The orange and purple stars are the $M_{\rm c,min}$ and $P_{\rm orb}$ for PSR\,J1932+2121 \citep{2024arXiv241203062W} and PSR\,J2129-0429 \citep{2024ApJ...966..161B}, respectively.}
\label{fig:orb_params}
\end{figure}

\begin{figure*}
\centering
\includegraphics[width=\linewidth]{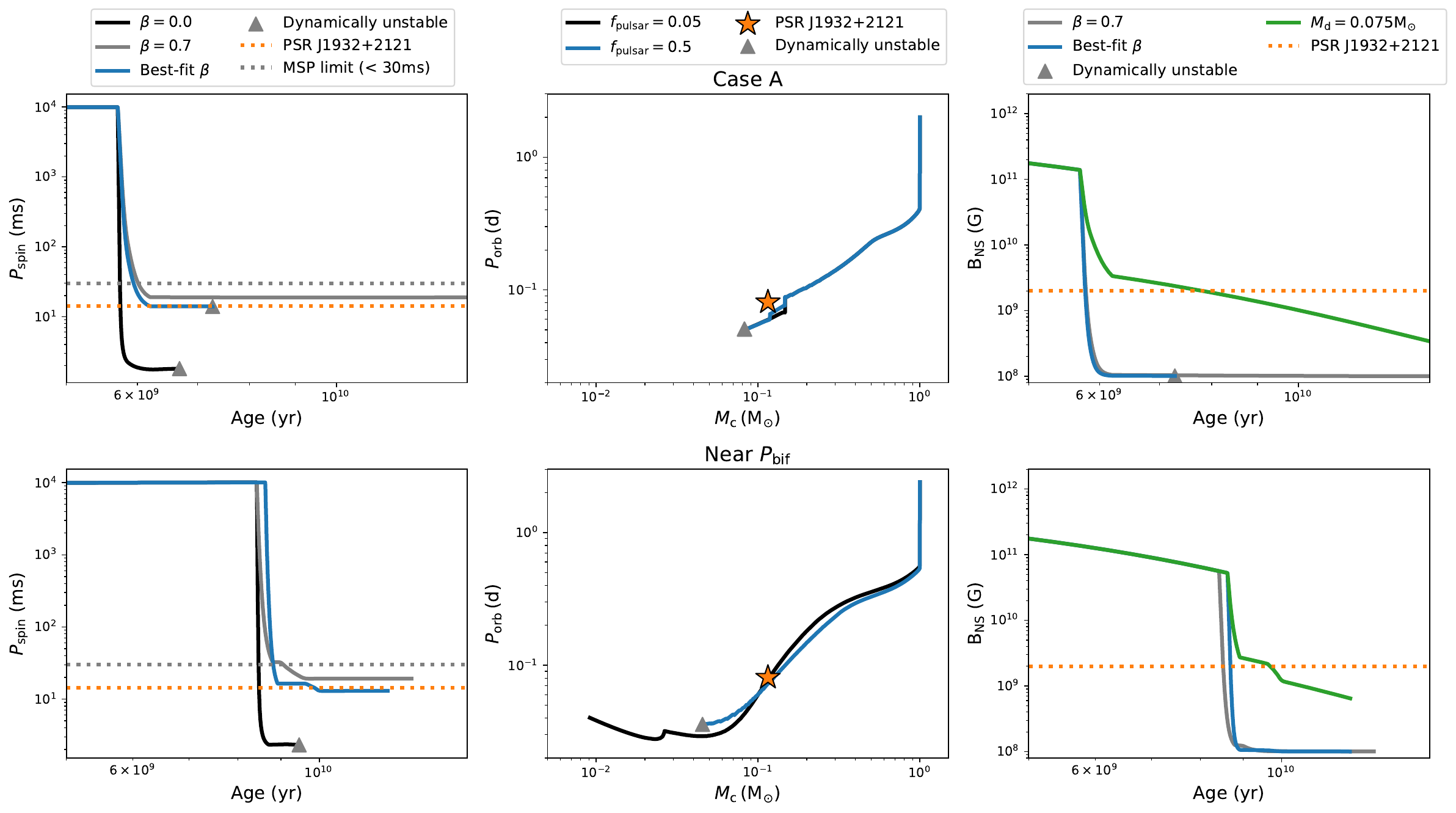}
\caption{Evolution of the case\,A (top row; with $P^{i}_{\rm orb}=2.0$) and near-$P_{\rm bif}$ (bottom row; $P^{i}_{\rm orb}=2.4$) binaries, for various physical assumptions. All binaries started with $M^{i}_{\rm NS}=1.3\,\Msun$ and $M^{i}_{\rm c}=1.0\,\Msun$. Left-most column shows $P_{\rm spin}$ versus age, for $\beta=0.0$ and 0.7 and the best-fit $\beta$ for both cases (0.67 for case\,A and 0.6 for near-$P_{\rm bif}$). Middle column shows $P_{\rm orb}$ versus $M_{\rm c}$, using the best-fitting $\beta$ value. Also, two values of $f_{\rm pulsar}$ (0.05 and 0.5) are shown for each RLO case. Right-most column shows the $B_{\rm NS}$ decay with time, for $\beta=0.7$ and the best-fit $\beta$. The value for $M_{\rm d}$ is 0.025\,\Msun. We also included a track for a higher $M_{\rm d}=0.075$\,\Msun, to study a slower decay of $B_{\rm NS}$. In all panels, the observed values from PSR J1932+2121 are marked in orange. }
\label{fig:compare}
\end{figure*}

LMXBs can be divided into three main types based on the evolutionary state of the companion at the onset of RLO: case\,A (main sequence), case\,B (companion has its exhausted core-H) and case\,C (later evolutionary stages). During case\,A RLO, binary orbits contract while for case\,B and C they expand. The orbital period separating cases\,A and B is the bifurcation period. Initial orbital periods just below the bifurcation period (near-$P_{\rm bif}$) have been associated with the formation of RBs and UCXBs \citep{2002ApJ...565.1107P,2003MNRAS.340.1214P,2019RAA....19..110H,2023ApJ...950...27G,2025A&A...693A.314M}. The evolutionary calculations presented in \citet{2025A&A...693A.314M}, show that case\,A and near-$P_{\rm bif}$ binaries can easily reproduce the observed BWs and RBs (including the $M_{\rm c,min}$ and $P_{\rm orb}$ of PSR\,J1932+2121). While values $\beta\lesssim0.3$ reproduce the typical spiders ($P_{\rm spin}\lesssim8$\,ms), higher values ($\beta\approx 0.7$) are needed for PSR\,J1932+2121 \citep[see figure\,2 in][]{2025A&A...693A.314M}.

Figure\,\ref{fig:orb_params} shows the orbital evolution of LMXBs that pass through the region where the binary parameters match those of PSR\,J1932+2121 (for $\beta=0.7$ and $f_{\rm pulsar}=0.05$) for both case\,A and near-$P_{\rm bif}$ binaries. We assume an initial NS mass of 1.3\,\Msun with initial orbital periods corresponding to case\,A RLO (2.0 and 2.2\,d) and near-$P_{\rm bif}$ RLO (2.4 and 2.6\,d). For the initial companion mass, we use $1.0$\,\Msun, since near-$P_{\rm bif}$ binaries that pass through the RB regime require $M^{i}_{\rm c}\lesssim1.5$\,\Msun \citep{2025A&A...693A.314M}. Case\,A binaries start RLO early in the companion main-sequence phase, whereas for near-$P_{\rm bif}$ binaries, the RLO phase takes place when the companion is close to the end of its main-sequence phase. As the RLO proceeds and the companion loses a significant amount of mass, it becomes fully convective causing magnetic braking stops to operate \citep{1983ApJ...275..713R}. For the case\,A binary, this occurs at $M_{\rm c}\sim0.15\,\Msun$ while for near-$P_{\rm bif}$ binaries, this happens at $M_{\rm c}\sim 0.027\,\Msun$ (these events are marked in Figure\,\ref{fig:orb_params}). This delayed turn-off of RLO is a direct consequence of the short duration of the RLO phase \citep{2025A&A...693A.314M}, which enables the system to pass through both the RB and UCXB regimes. 

In both cases of RLO, post-RLO evolution continues until the companion mass decreases to $M_{\rm c}\sim 0.01\,\Msun$ approaching the tidarren regime, and the evolution stops as the binary age reaches Hubble time. Only $P^{i}_{\rm orb}=2.2$\,d results in an unstable RLO. Irradiation removes material from the envelope of the fully-convective companion, that expands rapidly in response to even small amount of mass lost. This leads to a huge increase in the mass-transfer rate which can trigger a dynamical instability and start a common-envelope phase destroying the companion. Regardless, binaries with $M^{i}_{\rm c}=1.0\,\Msun$, $M^{i}_{\rm NS}=1.3\,\Msun$ and $P^{i}_{\rm orb}$ in the range of 2.0 to 2.6 are able to reproduce the observed parameters of the RB for a specific phase of their evolution. For initial orbital periods less than 2.0\,d the turn-off point moves to higher masses, resulting in overestimating the companion mass. For initial periods greater than 2.6\,d, the binary orbits widen during RLO instead of contracting. For $P^{i}_{\rm orb}$ in the range of 2.0 to 2.6\,d, the simulated companions masses at the observed orbital period of 0.08\,d are within 30\% of the observed minimum mass. In the following, we will focus on two cases: $P^{i}_{\rm orb}=2.0$ and 2.4\,d, to represent case\,A and near-$P_{\rm bif}$ binaries, respectively.

\begin{table}
\centering
\caption{Final NS spins $P^{f}_{\rm spin}$ for a range of $P^{i}_{\rm orb}$ and the corresponding $\beta$ values that were required to reproduce the observed value (14.245\,ms) within 10\% uncertainty.} 
\begin{tabular}{lcc}
\hline
$P^{i}_{\rm orb}$ (d) & $\beta$ & $P^{f}_{\rm spin}$ (ms) \\ \hline
2.0            & 0.67    & 14.17                                                   \\
2.2            & 0.68    & 13.66                                                   \\
2.4            & 0.6     & 13.04                                                   \\
2.6            & 0.5     & 13.1    \\ \hline                                      
\end{tabular}
\label{tab:beta_spin}
\end{table}


We constrain three aspects of binary evolution ($\beta$, $f_{\rm pulsar}$ and $B_{\rm NS}$ decay) in Figure\,\ref{fig:compare}, for case\,A (top row) and near-$P_{\rm bif}$ binaries (bottom row). In the left-most column, we show the spin evolution of case\,A and near-$P_{\rm bif}$ binaries, varying $\beta$ between 0.0 and 0.7, all with $f_{\rm pulsar}=0.05$. For fully conservative mass transfer, the final pulsar spins are 2.5\,ms (case\,A) and 2.7\,ms (near-$P_{\rm bif}$), in both cases the final spin is too fast for PSR\,J1932+2121. A highly non-conservative scenario ($\beta=0.7$) gives much slower spins, 19.0\,ms (case\,A) and 30.6\,ms (near-$P_{\rm bif}$). We fine tune $\beta$ to reproduce the observed spin of 14.245\,ms within 10\% uncertainty, with $\beta=0.67$ (case\,A) and 0.6 (near-$P_{\rm bif}$). The final NS masses are $1.58\,\Msun$ (case\,A) and $1.57\,\Msun$ (near-$P_{\rm bif}$), since the amount of accreted material is similar in both cases (about 0.2\,\Msun) and directly affects the final spins. The required value of $\beta$ is in the range of 0.5 to 0.67 for various values of $P^{i}_{\rm orb}$ (see Table\,\ref{tab:beta_spin}). Since lower values of $\beta$ (and faster final pulsar spins) lead to higher irradiation-induced mass loss, the RLO tends to be increasingly unstable.

Since PSR\,J1932+2121 shows signs of irradiation of its companion, as evidenced by radio eclipses covering about 8.6\% of the orbital phase \citep{2024arXiv241203062W}, we investigate varying values of $f_{\rm pulsar}$ (0.05 and 0.5) as shown in Figure\,\ref{fig:compare} (middle column). These binaries are simulated with the best-matching value of $\beta$ for the respective RLO case. For case\,A, increasing $f_{\rm pulsar}$ has no significant effect on the orbital evolution of the binaries, despite the pulsar having a spin-down luminosity comparable to typical spiders (see Table\,\ref{tab:J1932}). For near-$P_{\rm bif}$, binary interaction becomes dynamically unstable with $f_{\rm pulsar}=0.5$. Even if the companion is not fully convective when irradiation begins, as is the case for near-$P_{\rm bif}$, the evolved companion has convective regions in its envelope. Hence, in the case of strong irradiation, PSR\,J1932+2121 could evolve into an isolated pulsar having fully destroyed its companion. Due to the negligible effect of $f_{\rm pulsar}$ on pulsar spin and orbital periods around the observed region of PSR\,J1932+2121, it is not possible to determine the exact value of $f_{\rm pulsar}$.

The surface magnetic field strength of PSR\,J1932+2121 is estimated as $B_{\rm NS}=2.27\times10^{9}$\,G, which is an order of magnitude higher than that of other spiders and most MSPs. Figure\,\ref{fig:compare} (right most column) shows the evolution of the pulsar surface magnetic field as the binary transitions through detached and accretion phases, with $\beta=0.7$ and the best-fitting $\beta$ for the two RLO cases. All cases assume $M_{\rm d}=0.025$\,\Msun with an additional binary track presented for $M_{\rm d}=0.075$\,\Msun. With $M_{\rm d}=0.025$\,\Msun, irrespective of the accretion efficiency, the magnetic field decay due to accretion is relatively quick (within 300 to 400\,Myr from the onset of RLO) leading to final $B_{\rm NS}$ values of $\sim 10^8$\,G. With $M_{\rm d}=0.075$\,\Msun, the magnetic field decay aligns more closely with the observed estimate. In this case the final pulsar spin is 343\,ms (case\,A) and 174\,ms (near-$P_{\rm bif}$), an order of magnitude larger than the observed value. The slower magnetic field decay with $M_{\rm d}=0.075$\,\Msun, results in a larger magnetospheric radius during RLO, reducing the angular velocity of the transferred matter and leading to slower spin up {(see Equation\,\ref{eq:omega})}. Hence, there is a need for a better description of surface magnetic field decay in accreting pulsars, that can describe both the bulk of the MSP population and the ones with anomalous fields like in PSR\,J1932+2121.

\section{Discussion}
\label{sec:discussion}

\subsection{Mildly-recycled MSPs}
Along with PSR\,J1932+2121, there are many other mildly recycled MSP binaries, see Figure\,\ref{fig:Psdot_Ps_data}. Majority of these MSPs have CO or He WD companions. The formation channels for these pulsar+WD systems have been studied extensively \citep[for eg.,][]{2012MNRAS.425.1601T}. There are a few pulsars with ultra-light companions ($M_{\rm c,min}\sim0.02\,\Msun$) that have been suggested as spider candidates, like PSR\,J1727-2951 and PSR\,J1502-6752, with $P_{\rm spin}=28.4$ and 26.7\,ms, respectively \citep{2014MNRAS.439.1865N,2012MNRAS.419.1752K}. Both of them do not show eclipses. PSR\,1727-2951 is difficult to explain as a BW candidate since its $P_{\rm orb}=0.4$\,d \citep{2020MNRAS.493.1063C} implies strong pulsar wind irradiation, while its slow spin supports $\beta\sim0.7$ and would not drive strong pulsar winds \citep[see figures\,1 and B.1 in][]{2025A&A...693A.314M}. PSR\,J1502-6752 has $P_{\rm orb}=2.5$\,d \citep{2012MNRAS.419.1752K} which is at least a factor of 2 larger than any other BWs, even if strong irradiation effects are assumed. Slower pulsars with UL companions, J1744-3922 and B1831-00 ($P_{\rm spin}\gtrsim 150$\,ms) have been suggested to be a separate class resulting from highly magnetic pulsars, common-envelope evolution or accretion-induced collapse of WDs \citep{2007ApJ...661.1073B}. Hence, PSR\,J1932+2121 is the only confirmed mildly-recycled spider in the Galactic field.

\subsection{High surface magnetic fields: comparison with PSR\,J2129-0429}
The process of magnetic field decay in accreting pulsars is not well understood. Numerical simulations approximate this effect using an exponential decay (see Equation\,\ref{eq:B_accretion}). \citet{2020MNRAS.494.1587C} investigated $M_{\rm d}$ values raging from 0.01 to 0.05\,\Msun and found that values closer to 0.02\,\Msun best reproduced the observed properties of Galactic pulsars. For values less than 0.015\,\Msun, the magnetic field decay was enhanced, moving most pulsars below the pulsar death line. Conversely, $M_{\rm d}\gtrsim 0.05$\,\Msun failed to reproduce the observed spins. PSR\,J2129-0429 is another RB that lies at the higher tail-end of the spider spin distribution ($P_{\rm spin}=$ 7.61\,ms), and is another spider with an unexpectedly high surface magnetic field. Its estimated magnetic field strength is $1.58\times 10^{9}$\,G \citep[][also see Figure\,\ref{fig:Psdot_Ps_data}]{2024ApJ...966..161B}. In Figure\,\ref{fig:orb_params}, this RB lies in the more massive $M_{\rm c,min}$ and wider $P_{\rm orb}$ part of the parameter space \citep[$M_{\rm c,min}=0.4\,\Msun$ and $P_{\rm orb}=15.2$\,hrs;][]{2016ApJ...816...74B}. It is most likely at the start of its RLO phase, which explains its slow $P_{\rm spin}$ and high $B_{\rm NS}$, as the pulsar has not yet been recycled fully. PSR\,J1932+2121, however, has already undergone at least one RLO phase, making it peculiar since its $P_{\rm spin}$, orbital evolution and estimated $B_{\rm NS}$ cannot simultaneously be explained by current theoretical models.

\subsection{Accretion efficiency}
Several studies of LMXBs support $\beta>0.5$ \citep[assuming no other non-conservative effects;][]{1999A&A...350..928T,2001ASPC..229..423R,2002ApJ...565.1107P,  2011A&A...527A..83Z, 2012MNRAS.423.3316A, 2012MNRAS.425.1601T}, while other studies suggest that some MSPs require lower values of $\beta$ \citep[][]{2024MNRAS.tmp.2289K, 2025A&A...693A.314M}. For a pulsar to spin up to 14\,ms, it needs to accrete about $0.2$\,\Msun \citep[see figure\,11 in][]{2025A&A...693A.314M}. Our models suggest $\beta>0.5$ to explain the slow spin of PSR\,J1932+2121. Hence, PSR\,J1932+2121 may represent a case where the RLO process was disrupted, halting the pulsar spin up to values below 10\,ms. The inefficient RLO could be the result of the effects like the propeller phase or accretion disc instabilities.

\subsection{The Shklovskii correction}
An important caveat to be considered when interpreting $B_{\rm NS}$ estimated from observed $P_{\rm spin}$ and $\dot{P}_{\rm spin}$ values is the Shklovskii effect \citep{1970SvA....13..562S}, which accounts for the proper motion of the source. For sources with high proper motion and/or spin period, the intrinsic spin period derivatives could be overestimated if this effect is not accounted for.
Most MSPs have a proper motion of around 10\,mas\,yr$^{-1}$ \citep{2005AJ....129.1993M}, resulting in a correction factor of approximately $10\%$, which would not affect the $\dot{P}_{\rm spin}$ values significantly. 
For PSR\,J1932+2121, the proper motion is not well defined with current FAST constraints of $\mu_{\rm RA} = -11\pm12$\,mas/yr and $\mu_{\rm DEC} = -8\pm30$\,mas/yr (Z. L. Yang, priv. comm.), leading to a highly uncertain Shklovskii correction that could surpass the observed $\dot{P}_{\rm spin}$. Hence, the unusually high surface magnetic field could result from high proper motion of the source.

\subsection{Searching for the optical and infrared counterpart}

\begin{figure}
\centering
\includegraphics[width=\linewidth]{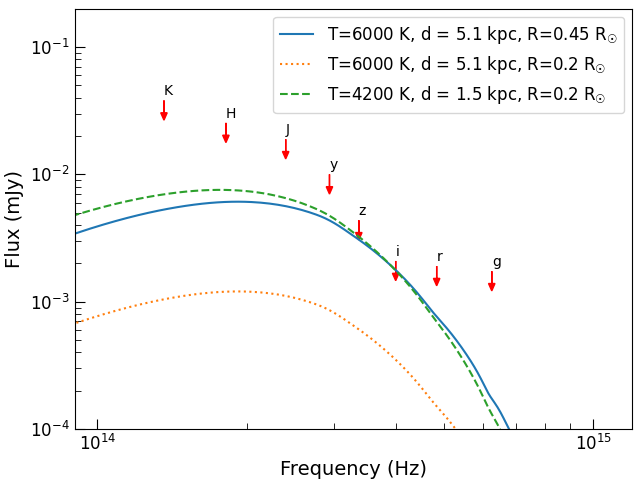}
\caption{The limiting fluxes from the UKIDSS survey (J-, H- and K-bands) and the Pan-STARRS survey ($y$, $z$, $i$, $r$ and $g$ bands) are indicated by arrows. The solid, dashed and dotted lines represent reddened blackbody spectra for various companion star parameter combinations, as shown in the legend.}
\label{fig:UL}
\end{figure}

We examined optical and infrared survey images at the location of PSR\,J1932+2121 using data from the Panoramic Survey Telescope and Rapid Response System (Pan-STARRS\footnote{https://catalogs.mast.stsci.edu/panstarrs/}), \textit{Gaia} DR3 \citep{gaia16b,gaia22k} and the United Kingdom Infrared Telescope Infrared Deep Sky Survey \citep[UKIDSS;][]{lawrence07,casali07}. No counterpart was found at the position of the FAST source. The nearest optical source identified in all above surveys is located approximately 1.06 arcseconds away (\textit{Gaia} position). Given the precise FAST position, an association with this star would require an unrealistically high proper motion ($>180$\,mas/yr). However, the \textit{Gaia} proper motion for this source is only 4.5\,mas/yr, effectively ruling out this possibility.

Using the limiting magnitudes of the Pan-STARRS and UKIDSS surveys ($g = 23.3$, $r = 23.2$, $i = 23.1$, $z = 22.3$, $y = 21.4$, $J = 20.7$, $H = 20.4$, $K = 20.0$), along with the estimated distance, extinction and stellar parameters derived from our synthetic binary evolution calculations, we can establish a lower limit on the distance and an upper limit on the companion star’s radius. \citet{2024arXiv241203062W} estimated the distance based on their measured dispersion measure (192.10\,cm$^{-3}$ pc) and the Galactic electron density models YMW16 \citep{yao17} and NE2001 \citep{cordes02}, obtaining values of 5.1\,kpc and 6.6\,kpc, respectively. Extinction is high in the Galactic plane; we used the 3D dust map \textsc{bayestar19} \citep{green19} to {estimate the extinction along the line of sight to PSR\,J1932+2121 at two distances: E(B-V) = 1.0 and 1.35 for distances of 1.5 and 5.1 kpc, respectively.}

Our synthetic binary evolution models suggest a companion temperature ($T$) of around 4200\,K and a stellar radius ($R$) of $0.2$\,$\rm R_\odot$ (which is also close to its Roche lobe radius). This temperature estimate assumes spherical symmetry and does not account for extra heating from pulsar wind irradiation, which likely increases the day-side temperature. RB pulsars typically exhibit day-side temperatures of approximately 6000\,K \citep{turchetta23,koljonen23}. In Fig. \ref{fig:UL}, we present blackbody spectra for several parameter combinations. The upper limits allow the modeled companion to remain undetected at 5.1 kpc. The survey limiting magnitudes set a lower distance limit of $d \gtrsim 1.5$\,kpc (for $T = 4200$\,K and $R = 0.2$\,$\rm R_\odot$) or an upper limit on the radius $ R \lesssim 0.45$\,$\rm R_\odot$ (for $T = 6000$ K and $d = 5.1$ kpc).

\section{Conclusions}
\label{sec:conclusions}

PSR\,J1932+2121 is a unique source since it is the only Galactic-field spider with a spin longer than 10\,ms. Typical spiders are better reproduced with more conservative accretion \citep{2025A&A...693A.314M}. Our analysis suggests that non-conservative accretion (where approximately 50 to 70\% of the mass transferred by the companion is lost, depending on the $P^{i}_{\rm orb}$) is required to reproduce the observed spin of PSR\,J1932+2121. Assuming an X-ray irradiated accretion disc and a $M^{i}_{\rm NS}=1.3\,\Msun$, the binary system that best matches the observed properties, has $M^{i}_{\rm c}=1.0\,\Msun$ and $P^{i}_{\rm orb}$ between 2.0 and 2.6\,d. This places the system close to its bifurcation period \citep{2025A&A...693A.314M}. Hence, PSR\,J1932+2121 serves as a valuable case study for understanding the causes and effects of inefficient mass accretion. The effect of pulsar wind irradiation is not constrained: a range of irradiation efficiencies explored (0.0 to 0.5) are able to reproduce the orbital properties of PSR\,J1932+2121. According to our evolutionary models, for a low level of pulsar wind irradiation, this source will likely evolve into a UCXB/tidarren.

The standard prescription describing accretion-induced surface magnetic field decay in pulsars is not able to reproduce the observed field strength of $2.27\times 10^{9}$\,G, underpredicting it by an order of magnitude. Varying the magnetic field mass decay scale in the model affects both the spin and the rate of $B_{\rm NS}$ decay, leading to much slower final pulsar spins ($\gtrsim 100$\,ms). However, the underlying physics of accretion-induced magnetic field decay is not well understood. Future timing observations estimating the proper motion of PSR\,J1932+2121 will be crucial in determining the Shklovskii correction leading to a more accurate estimate of the pulsar surface magnetic field, which can then be compared to magnetic field decay models on a more solid basis. One observable that could further constrain the binary initial conditions is the surface H abundance of the companion: case\,A companions would retain a higher surface H content than near-$P_{\rm bif}$ companions \citep{2025A&A...693A.314M}. Since the observed orbital parameters for PSR\,J1932+2121 can be reproduced for a range of irradiation strengths future studies combining optical light curves and atmospheric surface modeling of irradiated companions can provide more insight into the nature of irradiation \citep[eg.,][]{2024ApJ...973..121S}.

\section*{Acknowledgments}
This project has received funding from the European Research Council (ERC) under the European Union’s Horizon 2020 research and innovation programme (grant agreement No. 101002352, PI: M. Linares). We thank Z.\,L.\,Yang for sharing unpublished FAST limits on the proper motion of PSR J1932+2121.

The Pan-STARRS1 Surveys (PS1) and the PS1 public science archive have been made possible through contributions by the Institute for Astronomy, the University of Hawaii, the Pan-STARRS Project Office, the Max-Planck Society and its participating institutes, the Max Planck Institute for Astronomy, Heidelberg and the Max Planck Institute for Extraterrestrial Physics, Garching, The Johns Hopkins University, Durham University, the University of Edinburgh, the Queen's University Belfast, the Harvard-Smithsonian Center for Astrophysics, the Las Cumbres Observatory Global Telescope Network Incorporated, the National Central University of Taiwan, the Space Telescope Science Institute, the National Aeronautics and Space Administration under Grant No. NNX08AR22G issued through the Planetary Science Division of the NASA Science Mission Directorate, the National Science Foundation Grant No. AST-1238877, the University of Maryland, Eotvos Lorand University (ELTE), the Los Alamos National Laboratory and the Gordon and Betty Moore Foundation.

\section*{Data Availability}

The \mesa\,inlists used in this article will be shared on reasonable request to the corresponding author.



\bibliographystyle{mnras}
\bibliography{main} 



\appendix




\bsp	
\label{lastpage}
\end{document}